\documentclass{iau}
\usepackage{graphicx}
\title[p-40 (s-05) GRB as luminosity indicator] 
{GRB as luminosity indicator}

\author[Basak, R. \& Rao, A. R.]   
{Rupal Basak
 \and A. R. Rao}

\affiliation{Tata Institute of Fundamental Research, Mumbai}

\pubyear{2013}
\volume{296}  
\pagerange{119--126}
\jname{Supernova environmental impacts}
\editors{Editors: Dick McCray \& Alak Ray}
\begin{document}

\maketitle

\begin{abstract}
Gamma Ray Bursts (GRBs) are found at much higher redshifts (z$>$6) than Supernova Ia (z$\sim$1), and hence, they can be used to probe 
very primitive universe. However, radiation mechanism of GRB remains a puzzle, unlike Supernova Ia. Through comprehensive description,
both empirical and physical, we shall discuss the most likely way to use the constituent pulses of a GRB to find the radiation mechanism 
as well as using the pulses as luminosity indicators.

\end{abstract}

\firstsection 

\section{Introduction}
First discovered during late 1960's, Gamma Ray Burst (GRB) soon became one of the greatest puzzles in astrophysics in terms of its location, size
and energetics. Many satellites have been flown since then to understand this puzzling phenomenon --- \emph{HETE-2}, \emph{BATSE}, \emph{Swift}, 
\emph{Fermi}, to name a few. Though we have a clear idea about their cosmic origin (the highest spectroscopic redshift being 8.2 for GRB 090423)
and a rough idea about their energy budget (highest among all astrophysical phenomena --- $\sim 10^{52}$ erg), the emission mechanism
is still unsettled. Hence, in spite of a great hope that GRB could be used as luminosity indicator in extension to the currently
used ones, e.g., supernova Ia, one has to standardize the GRB energetics first. This, of course, serves two purposes --- solving the 
GRB physics, which is not settled till now, and using GRB as luminosity indicator.

\section{Methodology and Results}
There exist certain empirical correlations of the peak energy ($E_{peak}$) of GRB spectrum with the energetics of GRB. These correlations 
are important as they can be used to independently measure a physical parameter, namely energy, using only prompt emission spectral
data. Amati et al. (2002) showed that $E_{peak}$ correlates with the isotropic equivalent energy ($E_{\gamma,iso}$). It is very important 
that this correlation should hold within a GRB, as that can prove the reality of such correlation and strongly refutes selection bias. But, 
Basak \& Rao (2012b), using 9 GRBs with known redshift detected by \emph{Fermi}/Gamma Ray Burst Monitor (GBM), have shown that this 
correlation breaks down if one uses the 
time-resolved data. The Pearson correlation, 0.80 drops to 0.37. They concluded that Amati correlation has no meaning in a 
time-resolved study. The situation is saved if one uses the constituent broad pulses (total 22 pulses), rather than intensity guided 
time cuts. Pulse-wise analysis not only restores the correlation, it improves that (0.89). They used the pulse description of 
Basak \& Rao (2012a) and found that replacement of $E_{peak}$ with a new quantity of their model, namely the peak energy at 
zero fluence ($E_{peak,0}$) improves the correlation even further (0.96).

In this study, we have enlarged our sample to 19 GRBs (43 pulses) having measured redshifts (z) from GBM catalog. We consider 
only pulse-wise Amati correlation here. The Spearman rank correlation coefficient is 0.86 (see Figure 1). As we have a larger sample, we are able to 
divide the set of GRBs into various redshift (z) bins and study evolution of the correlation. A detailed analysis will be published later.

\begin{figure}\centering
 \includegraphics[width=4.0 in]{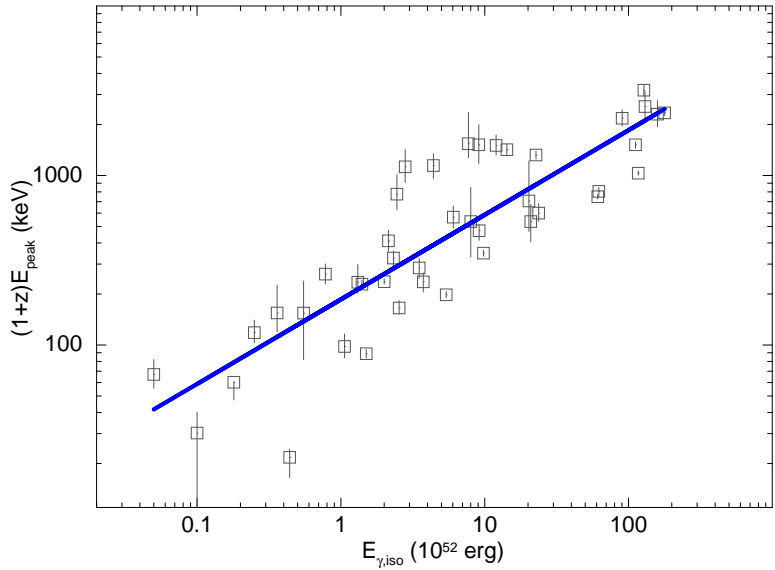}
     
\caption{Pulse-wise Amati correlation. The thick line shows the correlation}
\end{figure}

\section{Discussions}
The prompt emission spectrum of a GRB is generally fitted with Band model (Band et al. 1993). The empirical correlations, we have 
discussed, relies on the fact that the overall and instantaneous spectrum is Band like. Band model describes a non-thermal model.
There are alternative models, e.g., black body with a powerlaw (Ryde 2004).
Recently, we have analyzed the brightest 
GRBs, having separable pulses, namely GRB 081221 and GRB 090618, and found that the Band model is adequate in the falling
part of a pulse. But, a different model is preferred in the rising part (Basak \& Rao 2013). Hence, one should 
take into account these findings while describing the emission mechanism of GRB and thereby using the pulses for cosmological purpose.


\begin{thebibliography}{}


\bibitem[Amati, Frontera, Tavani \etal\ (2002)]{Amati02}
{Amati, L., Frontera, F., Tavani, M., in't Zand, J. J. M., Antonelli, A., Costa, E., Feroci, M., Guidorzi, C., Heise, J., Masetti, N., Montanari, E., Nicastro, L., Palazzi, E., Pian, E., Piro, L., Soffitta, P.} 2002, 
\textit{A\&A}, 390, 81

\bibitem[Band, Matteson, Ford, \etal\ (1993)]{Band93}
{Band, D., Matteson, J., Ford, L., Schaefer, B., Palmer, D., Teegarden, B., Cline, T., Briggs, M., Paciesas, W., Pendleton, G., Fishman, G., Kouveliotou, C., Meegan, C., Wilson, R., Lestrade, P.} 1993, 
\textit{ApJ}, 413, 281

\bibitem[Basak, Rao (2012a)]{Basak12a}
{Basak, R., Rao, A. R.} 2012a, 
\textit{ApJ}, 745, 76

\bibitem[Basak, Rao (2012b)]{Basak12b}
{Basak, R., Rao, A. R.} 2012b, 
\textit{ApJ}, 749, 132

\bibitem[Basak, Rao (2013)]{Basak13}
{Basak, R., Rao, A. R.} 2013, 
\textit{arXiv}: 1302.6091


\bibitem[Ryde (2004)]{Ryde04}
{Ryde, F.} 2004, 
\textit{ApJ}, 614, 827





\end{thebibliography}
\end{document}